# Evolving Boolean Regulatory Networks with Variable Gene Expression Times

Larry Bull

Department of Computer Science & Creative Technologies,
University of the West of England, Bristol BS16 1QY, U.K.

Larry.Bull@uwe.ac.uk

**Abstract.** The time taken for gene expression varies not least because proteins vary in length considerably. This paper uses an abstract, tuneable Boolean regulatory network model to explore gene expression time variation. In particular, it is shown how non-uniform expression times can emerge under certain conditions through simulated evolution. That is, gene expression time variance can be beneficial in the shaping of the dynamical behaviour of the regulatory network without explicit consideration of protein function.

## 1. Introduction

A protein's function is dependent upon its tertiary (3D) structure which in turn is dependent upon the primary structure of the amino acid sequence by which it is specified. Typically, the more amino acids in the primary structure, the more complex the tertiary structure. Similarly, the more amino acids, the longer gene expression can be expected to take. The lengths of genes/amino acid sequences varies considerably within and across taxa. It is well-established that, due to chemical equivalences between amino acid sequences, there is a strong neutrality effect at the molecular level of evolution (eg, [11]). However, this does not fully explain the differences in the distribution of protein lengths seen, nor why eukaryotic proteins are typically larger than bacterial proteins (eg, [15]).

With the aim of enabling the systematic exploration of artificial genetic regulatory network models (GRN), a simple approach to combining them with abstract fitness landscapes has been presented [2]. More specifically, random Boolean networks (RBN) [8] were combined with the NK model of fitness landscapes [10]. In the combined form – termed the RBNK model – a simple relationship between the states of $N$ randomly assigned nodes within an RBN is assumed such that their value is used within a given NK fitness landscape of trait dependencies. This paper explores the introduction of variable expression times to the genes in a traditional Boolean regulatory network within the RBNK model. That is, the effects of protein length variation are considered based purely upon the dynamical behaviour of the regulatory network being shaped under an evolutionary process. It is shown that non-uniform gene expression times can be selected for and that a relationship appears to exist between gene length and gene connectivity in such cases.

## 2. The RBNK Model

Within the traditional form of RBN, a network of $R$ nodes, each with a randomly assigned Boolean update function and $B$ directed connections randomly assigned from other nodes in the network, all update synchronously based upon the current state of those $B$ nodes (Figure 1). Hence those $B$ nodes are seen to have a regulatory effect upon the given node, specified by the given Boolean function attributed to it. Since they have a finite number of possible states and they are deterministic, such networks eventually fall into an attractor. It is well-established that the value of $B$ affects the emergent behaviour of RBN wherein attractors typically contain an increasing number of states with increasing $B$ (see [9] for an overview). Three regimes of behaviour exist: ordered when $B$=1, with attractors consisting of one or a few states; chaotic when $B\geq3$, with a very large number of states per attractor; and, a critical regime around $B$=2, where similar states lie on trajectories that tend to neither diverge nor converge (see [3] for formal analysis). Note that traditionally the size of an RBN is labelled $N$, as opposed to $R$ here, and the degree of node connectivity labelled $K$, as opposed to $B$ here. The change is adopted due to the traditional use of the labels $N$ and $K$ in the NK model of fitness landscapes which are also used in this paper, as will be shown.

Kauffman and Levin [10] introduced the NK model to allow the systematic study of various aspects of fitness landscapes (see [9] for an overview). In the standard NK model an individual is represented by a set of $N$ (binary) genes or traits, each of which depends upon its own value and that of $K$ randomly chosen others in the individual (Figure 1). Thus increasing $K$, with respect to $N$, increases the epistasis. This increases the ruggedness of the fitness landscapes by increasing the number of fitness peaks. The NK model assumes all epistatic interactions are so complex that it is only appropriate to assign (uniform) random values to their effects on fitness. Therefore for each of the possible $K$ interactions, a table of $2^{(K+1)}$ fitnesses is created, with all entries in the range 0.0 to 1.0, such that there is one fitness value for each combination of traits. The fitness contribution of each trait is found from its individual table. These fitnesses are then summed and normalised by $N$ to give the selective fitness of the individual. Exhaustive search of NK landscapes [14] suggests three general classes exist: unimodal when $K$=0; uncorrelated, multi-peaked when $K$>3; and, a critical regime around $0<K<4$, where multiple peaks are correlated.

As shown in Figure 2, in the RBNK model $N$ nodes (where $0<N\leq R$) in the RBN are chosen as outputs, ie, their state determines fitness using the NK model. The combination of the RBN and NK model enables a systematic exploration of the relationship between phenotypic traits and the genetic regulatory network by which they are produced. It was previously shown how achievable fitness decreases with increasing $B$, how increasing $N$ with respect to $R$ decreases achievable fitness, and how $R$ can be decreased without detriment to achievable fitness for low $B$ [2]. In this paper $N$ phenotypic traits are attributed to randomly chosen nodes within the network of $R$ genetic loci (Figure 2). Hence the NK element creates a tuneable component to the overall fitness landscape. Self-connection by nodes is allowed.

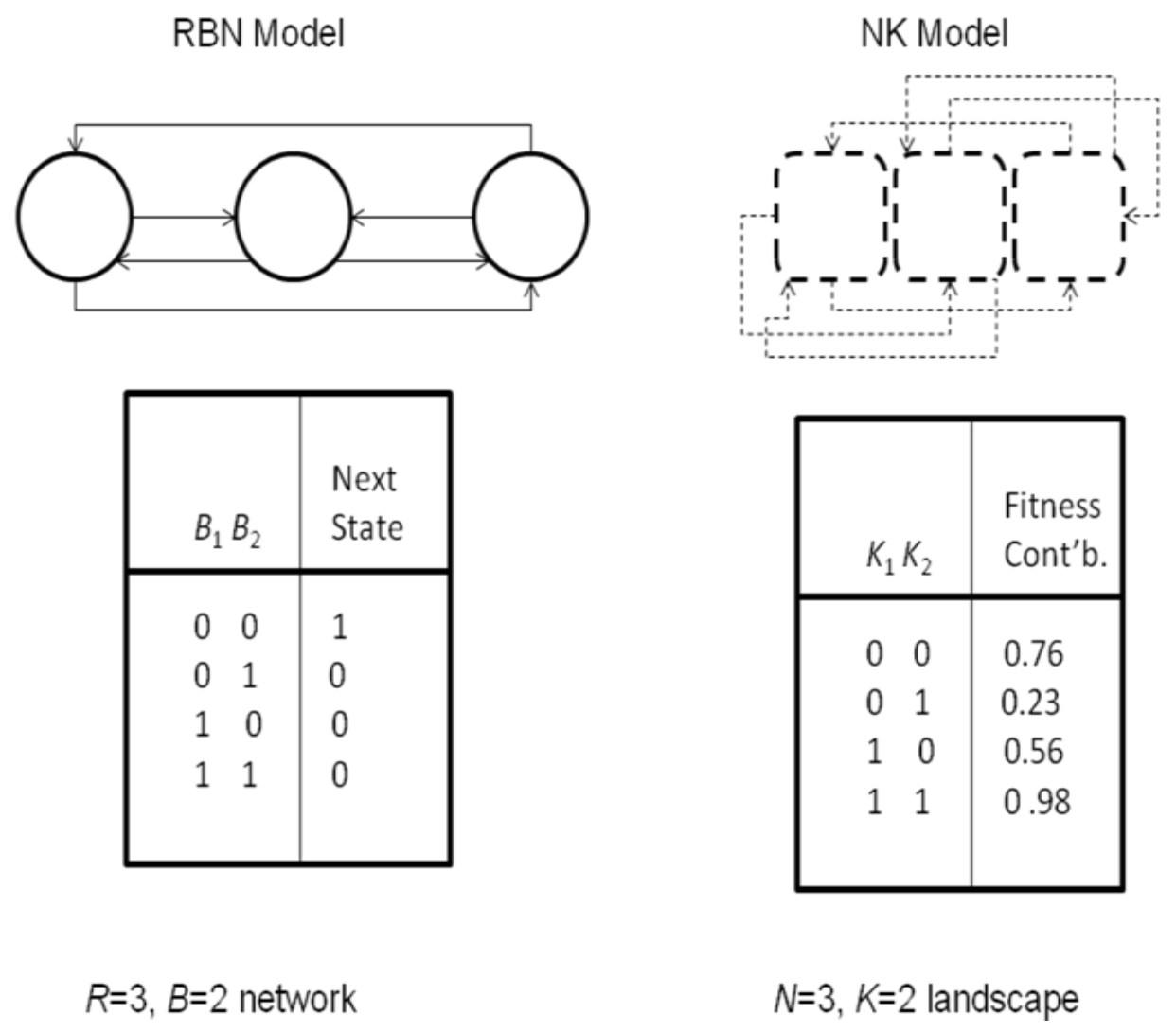

**Fig.1.** Example traditional RBN (left) and NK (right) models. Both contain three genes mutually connected, with the state-transition/fitness-contribution table shown for one gene in each case.

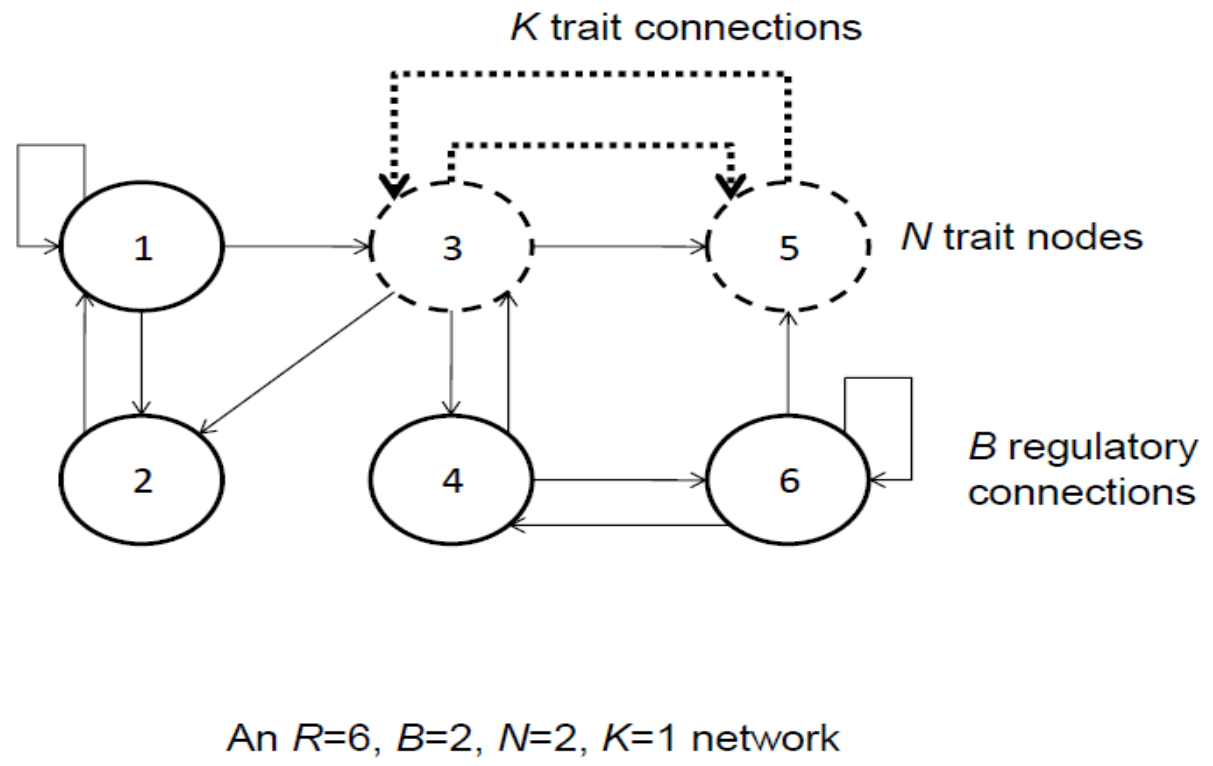

**Fig. 2.** Example RBNK model. Dashed lines and nodes indicate where the NK fitness landscape is embedded into the RBN model.

# 3. The RBNK Model with Variable Gene Expression Times

## 3.1 Gene Expression

Within the traditional form of RBN each node updates synchronously, in parallel, taking one time-step. That is, each gene has the same length or expression time: one update cycle. To include a mechanism which enables the variation in time taken for expression, each node in the RBN is extended to (potentially) include a time from a specified range by which to delay updating its state once turned on. Hence on each cycle, each single time-step node updates its state based upon the current state of the *B* nodes it is connected to using the Boolean logic function assigned to it in the standard way. Any nodes marked as having a longer expression time are checked to see if they have previously been switched on and have waited the number of update cycles associated with them. If this is the case, the node is either turned on and the counter reset, otherwise its counter is incremented and it remains off. Such nodes remain on until they are switched off by their Boolean function; returning to the off state and a subsequent delay for continued expression is not explored here.

## 3.2 Experimentation

For simplicity with respect to the underlying evolutionary search process, a genetic hill-climber is considered here, as in [2]. Each RBN is represented as a list to define each node's Boolean function, *B* connection ids, an update time delay, and whether it is a time delayed node or not. Mutation can therefore either (with equal probability): alter the Boolean function of a randomly chosen node; alter a randomly chosen *B* connection; turn a node into or out of being a time delayed node; or, alter a time delay, if it is a delayed node. A single fitness evaluation of a given GRN is ascertained by updating each node for 100 cycles from a randomly defined genome start state. On the last update cycle, the value of each of the *N* trait nodes in the GRN is used to calculate fitness on the given NK landscape. This process is repeated ten times per run. A mutated GRN becomes the parent for the next generation if its fitness is higher than that of the original. In the case of fitness ties the number of time delayed nodes is considered, with the smaller number favoured, the decision being arbitrary upon a further tie. *Hence there is a slight selective pressure against variable expression times*. Here $R$=100, $N$=10 and results are averaged over 100 runs - 10 runs (each of 10 random starts) on each of 10 landscapes per parameter configuration - for 5000 generations. As in [2], $0<B\leq 5$ and $0\leq K\leq 5$ are used. Expressions times ($T$) were able to vary up to 10 update cycles ($1\leq T\leq 10$), where $T$=0 is the traditional case of no delay.

Figure 3 (left column) shows how there is a significant (T-test, $p<0.05$) drop in fitness for $B>2$ compared to $B<3$ regardless of $K$. Moreover, it can be seen that around 3% of nodes, on average, have an expression time of longer than one update cycle for $B>1$, for all $K$. Figure 3 (right column) also shows the average expression time for those nodes. As can be seen, for $B$=2, $T\approx 2$, whereas there appears to be no selective pressure on $T$ in the cases where evolution struggles to produce high fitness networks, ie, when $B>2$, with average times of around 5 cycles (in a range 1-10) typically seen.

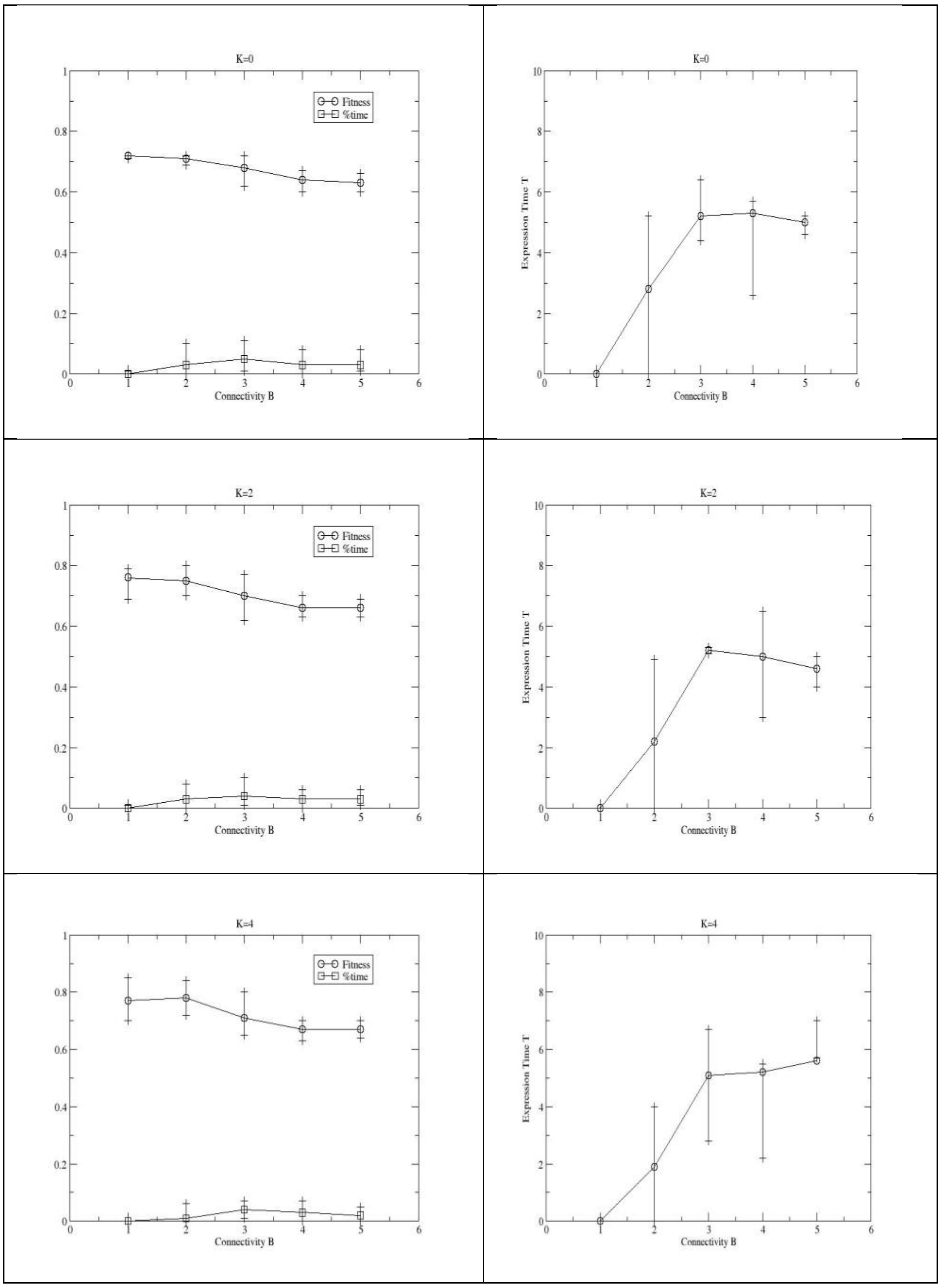

**Fig. 3.** Evolutionary behaviour after 5000 generations in the RBNK model for *R*=100, *N*=10 and various *B* and *K* combinations. Left column shows fitnesses and the percentage of nodes with delayed expression, the right column shows the corresponding average delay. Error bars show min and max values.

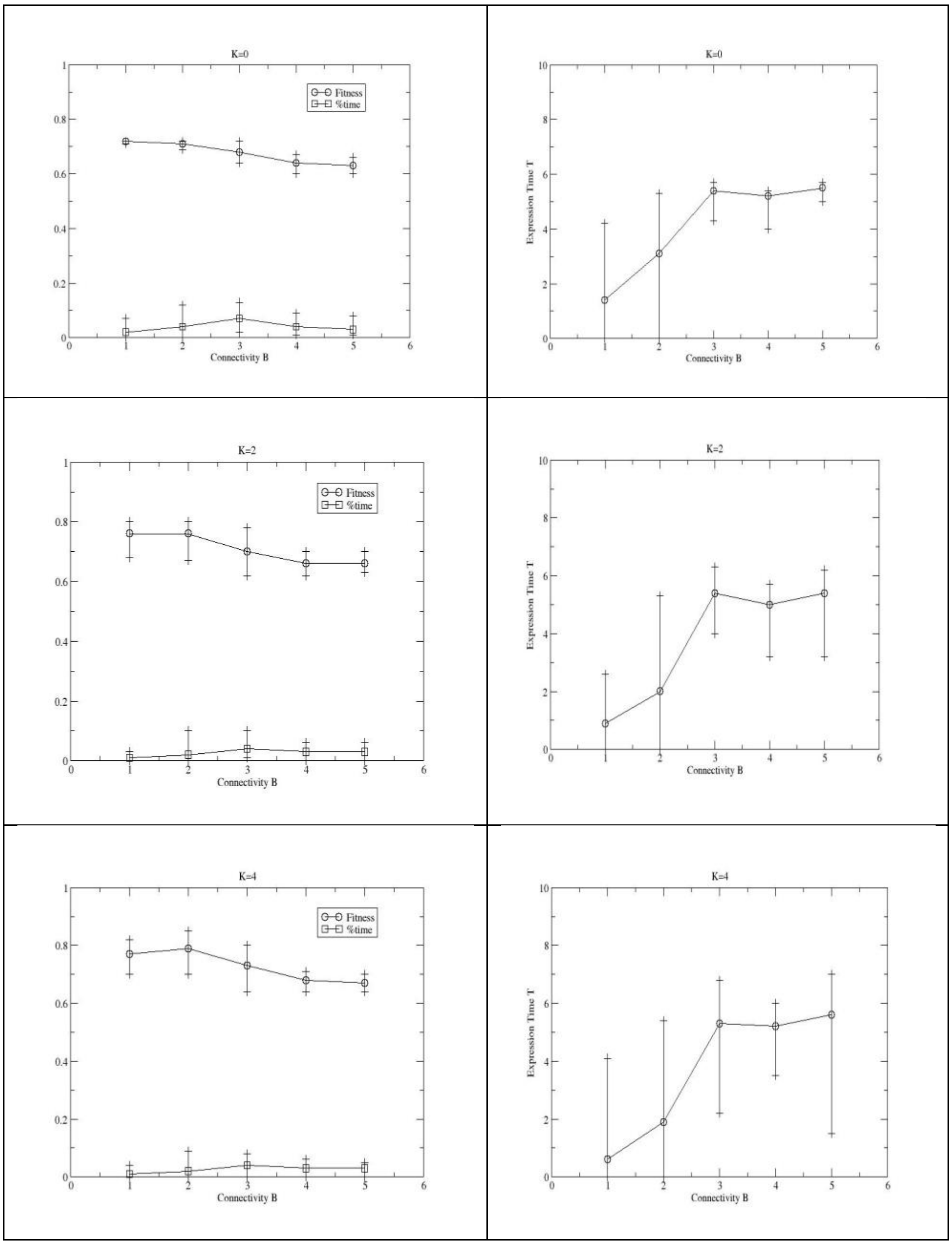

**Fig. 4.** Evolutionary behaviour after 5000 generations in the asynchronous RBNK model for *R*=100, *N*=10 and various *B* and *K* combinations. Left column shows fitnesses and the percentage of nodes with delayed expression, the right column shows the corresponding average delay. Error bars show min and max values.

### 3.3 Asynchronous Experimentation

As noted above, traditional RBN update synchronously, ie, a global clock signal is assumed to exist. It has long been suggested that this assumption is less than realistic for natural systems and hence discrete dynamical models have also used asynchronous updating (eg, [13]). Harvey and Bossomaier [7] presented an asynchronous form of RBN wherein a node is picked at random (with replacement) to be updated, with the process repeated *R* times per cycle to give equivalence to the synchronous case. The resulting loss of determinism means such networks no longer fall into regular cyclic attractors, rather they typically fall into so-called "loose" attractors where "the network passes indefinitely through a subset of its possible states" [13]. Many forms of asynchronous updating are possible (eg, see [5] for an overview) but the simple random scheme is used here. It can also be noted that Gershenson [4] has introduced a deterministic asynchronous update scheme wherein individual nodes are given an update frequency from a range, similar to above. Using very small RBN, it is suggested that such updating typically increases the length and number of attractors for low network connectivity *B* [5].

Figure 4 (left column) shows how there is a significant (T-test, $p<0.05$) drop in fitness for $B>2$ compared to $B<3$ regardless of *K*, as in the synchronous case above (see also [2]). Again, it can be seen that around 3% of nodes, on average, have an expression time of longer than one update cycle for all *K*. In contrast to the synchronous case above, this is also true for $B=1$. As noted in Section 2, such RBN typically exhibit point attractors whereas this is not the case for asynchronous updating. Analysis in all cases shows that the delayed nodes form part the subset changing state within the attractor, which corresponds with the $B=1$ result for synchronous updating as they are not used. Figure 4 (right column) shows the average expression time for the delayed nodes. As can be seen, there again appears to be no selective pressure on *T* in the cases where evolution struggles to produce high fitness networks, ie, when $B>2$. For $B=1$, $T\approx1$ which is significantly (T-test, $p<0.05$) less than for $B=2$ where $T\approx2$. The latter value for *T* is the same as for synchronous updating above. Similar general findings were found for other ranges of *T* (not shown).

## 4. Variable Sized GRN with Variable Gene Expression Times

### 4.1 Emergent Complexity

In the above experimentation, the total number of nodes *R* within the GRN was fixed. Using a version of the NK model, Harvey [6] showed, by including a bias, that gradual growth through small increases in genome length via mutation is sustainable whereas large increases in genome length per growth event is not sustainable. This is explained as being due to the fact that a degree of correlation between the smaller fitness landscape and the larger one must be maintained; a fit solution in the former space must achieve a suitable level of fitness in the latter to survive into succeeding generations. Kauffman and Levin [10] discussed this general concept with respect to fixed-size NK landscapes and varying mutation step sizes therein. They showed how for long jump adaptations, ie, mutation steps of a size which go beyond the correlation length of a given fitness landscape, the time taken to find fitter variants doubles per generation. Harvey's [6] growth operator is a form of mutation which adds *g* random genes to an original genome of length *G*. Hence he draws a direct analogy between the size of *g* and the length of a jump in a traditional landscape; the larger *g*, the less correlated the two landscapes will be regardless of the underlying degree of correlation of each. Aldana et al. [1] have examined the effects of adding a new, single gene into a given RBN through duplication and divergence. They find, somewhat reminiscent of Harvey's result, that the addition of one gene typically only slightly alters the attractors of the resulting RBN when $B$<3. Attractor structure is not conserved in the chaotic regime, however.

The experiments reported above have been repeated with the addition of two extra "macro" mutation operators: one to delete the end node (the *N* trait nodes cannot be deleted), randomly re-assigning any connections to it; and, one to add a random node on to the end of the genome, connecting it to a randomly chosen node in the network. These two operators occur with equal probability to the previously described mutation operators. The replacement process is also altered such that, when fitnesses and the number of delayed nodes are equal, the smaller network is kept, with ties again broken at random. RBNs were initialised with $R$=100, as before.

### 4.2 Experimentation

In all cases, no significant change in the fitness of solutions is seen (not shown). There is also, typically, no significant effect on the resulting size of the networks. However, as can be seen in Figure 5, for low connectivity ($B$<3), regardless of *K* and the updating scheme, the networks decrease in size by around a half – a statistically significant change (T-test, $p<0.05$). That is, not only do low connectivity networks evolve the highest fitnesses for all *K*, they are able to do so with a *smaller number* of nodes *R*. It is known that both the number of states in an attractor and the number of attractors are dependent upon *R* within RBN, and that the general form of those relationships changes for low and high connectivity. For example, with synchronous updating, when $B$=2, attractors are typically of size $R^{0.5}$, whereas, when $B$=$R$, attractors typically contain 0.5 x $2^{R/2}$ states (eg, see [9] for a summary). Hence the evolutionary process appears able to exploit the potential for ever smaller attractors for the low *B* cases, driven by the selection pressure for network size reduction, and to do so whilst maintaining fitness. In this case, incrementally decreasing *R* is sustainable since the subsequent change in the attractors is sufficiently small for low *B*, whereas the same change in *R* appears to cause a significant change in the attractor space for high *B*. This result is somewhat anticipated by those of Aldana et al. [1] and Harvey [6] but is also in the opposite direction: small *reductional* changes are maintained. This general result was also found in [2] but is slightly altered by the inclusion of an expression delay since the networks for $B$=1 are larger here for both update schemes (not shown). That is, as the results in Section 3 suggest, varying the gene expression time provides evolution with a mechanism through which to alter the attractor space. Varying the size of the network is another mechanism through which this can be achieved. When both mechanisms are made available to evolution, both appear to be used; there is a trend towards fewer delayed nodes per network when size is also varying. It can also be noted that a drop in the average delay for *B*=2 with asynchronous updating is seen when size is varying.

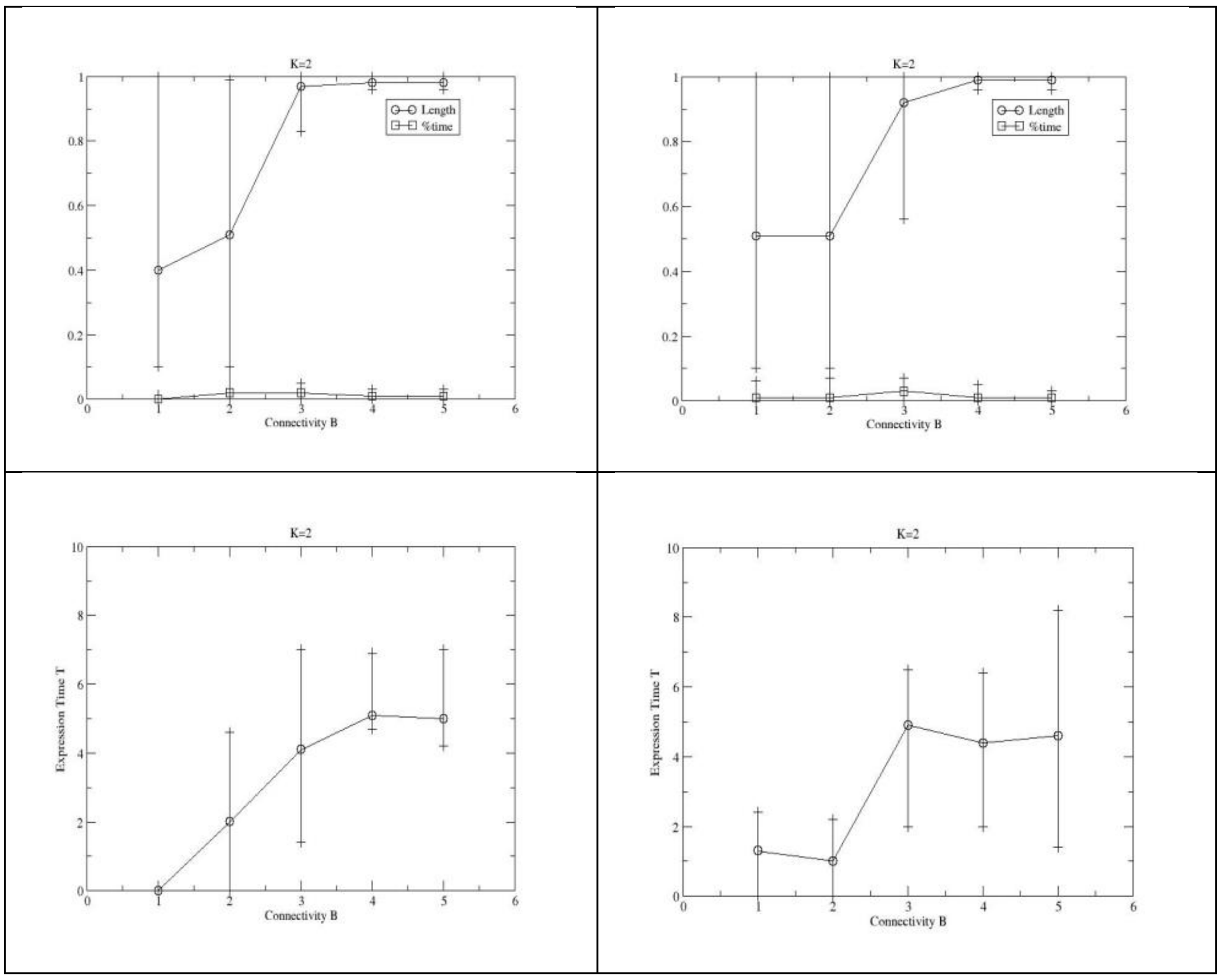

**Fig. 5.** Example behaviour after 5000 generations when the RBN size is able to vary. Left column shows results for synchronous updating, with the fraction of the original length and the percentage of nodes with delayed expression (top), and the average expression time (bottom). The right column shows the same for asynchronous updating.

## 5. Conclusions

The length of genes and hence the length of the amino acid sequences which specify proteins varies considerably within and across taxa. This paper has considered the effects of protein length variation based purely upon the dynamical behaviour of regulatory networks being shaped under evolution. It has been shown that non-uniform gene expression times can be selected for under low gene connectivity and that *such genes typically have an expression delay proportional to the degree of connectivity*. This general result appears to correspond with the natural case: it is known that eukaryotic organisms typically exhibit longer proteins than prokaryotes (eg, [15]); and, it is also known that organisms such as *e.coli* have a lower average level of gene connectivity than higher organisms such as *S.cerevisiae* (eg, [12]). As such, the use of non-uniform gene expression times may prove beneficial in work considering the evolution of artificial genetic regulatory networks (eg, see [2] for an overview) for computation.